\documentclass[aps,prd,twocolumn,showpacs,superscriptaddress]{revtex4}  
\usepackage{times}
\usepackage{graphicx}
\usepackage{dcolumn}   
\usepackage{bm}        
\usepackage{amssymb}   
\usepackage{amsmath}
\usepackage{xspace}

\begin{document}

\title{Stochastic Optimization for Collision Selection in High Energy Physics}

\author{S. Whiteson}
\affiliation{Dept. of Computer Science, Unverisity of Texas, Austin, Texas}

\author{D. Whiteson}
\affiliation{Dept. of Physics and Astronomy, University of Pennsylvania, Philadelphia, Pennsylvania}

\begin{abstract}
  The underlying structure of matter can be deeply probed via
  precision measurements of the mass of the \emph{top quark}, the most
  massive observed fundamental particle.  Top quarks can be produced
  and studied only in collisions at high energy particle accelerators.
  Most collisions, however, do not produce top quarks; making precise
  measurements requires culling these collisions into a sample that is
  rich in collisions producing top quarks (\emph{signal}) and spare in
  collisions producing other particles (\emph{background}).  Collision
  selection is typically performed with heuristics or supervised
  learning methods.  However, such approaches are suboptimal because
  they assume that the selector with the highest classification
  accuracy will yield a mass measurement with the smallest statistical
  uncertainty.  In practice, however, the mass measurement is more
  sensitive to some backgrounds than others.  Hence, this paper
  presents a new approach that uses stochastic optimization techniques
  to directly search for selectors that minimize statistical
  uncertainty in the top quark mass measurement.  Empirical results
  confirm that stochastically optimized selectors have much smaller
  uncertainty.  This new approach contributes substantially to our
  knowledge of the top quark's mass, as the new selectors are
  currently in use selecting real collisions.
\end{abstract}

\maketitle

\section{Introduction}
\label{sec:introduction}

The underlying structure of matter and the laws that govern its
interaction remain compelling mysteries.  Physicists hope to solve
these mysteries with the help of modern high energy accelerators,
which collide protons and anti-protons to create exotic particles that
have not existed since the early universe.  Of particular interest is
the \emph{top quark}, the most massive observed fundamental particle
and nearly as massive as a gold atom.  The top quark is intriguing not
only because of its mass, but because of what it may reveal about the
nature of mass itself: precision measurements of its mass stringently
test theories that attempt to explain the origins of particle
mass~\cite{toprole2,toprole1,topsusy,higgs}.

Only the world's most powerful collider, the FermiLab Tevatron in
Batavia, Illinois, has sufficient energy to produce top
quarks~\cite{topdiscovery1,topdiscovery2}.  Even so, out of
approximately $10^{10}$ collisions per hour, on average fewer than one
produces a top quark.  Since collisions are extraordinarily expensive
to generate, maximizing the precision of the resulting mass
measurement is critical. Doing so requires culling these collisions
into a sample that is rich in collisions producing top quarks
(\emph{signal}) and spare in collisions producing other particles
(\emph{background}).  Collision selection is difficult because several
types of background mimic the top quark's characteristic signature.

However, these difficulties can be overcome with the help of machine
learning.  Previous research on related collision selection problems
used \emph{supervised learning} methods to train neural
networks~\cite{nn2,nn1} or support vector machines~\cite{svm} that
classify collisions as signal or background.  While this approach has
proven effective, it is applicable only to the narrow class of
problems where higher classification accuracy consistently yields more
precise measurements.  The measurement of the top quark mass
exemplifies a broader class of problems where this assumption does not
hold. Instead, the mass measurement is more sensitive to the presence
of some background collisions than others, in ways that are difficult
to predict \emph{a priori}.  Therefore, selectors that maximize
classification accuracy may perform worse than those that 1) increase
the quantity of signal by tolerating harmless background or 2) reduce
the quantity of signal to eliminate disruptive background.

This paper presents a new approach that uses \emph{stochastic
  optimization} techniques to find such a selector.  Rather than
maximizing classification accuracy, this approach directly searches
for selectors that yield mass measurements with the smallest
statistical uncertainty.  Using NEAT~\cite{neat}, an evolutionary
method for training neural networks, we train collision selectors that
operate either in conjunction with supervised classifiers or in lieu
of them.

We present experiments that compare the performance of neural network
selectors trained with backpropagation~\cite{backprop}, a supervised
method, to those trained with NEAT.  The results confirm the advantage
of the stochastic optimization approach, as the NEAT selectors yield
much more precise mass measurements.  These NEAT selectors are
currently in use at FermiLab for selecting collisions from real data
collected with the Tevatron collider.  Hence, this new approach to
collision selection contributes substantially to our knowledge of
the top quark's mass and our understanding of the larger questions
upon which it sheds light.

\section{Measuring the Top Quark's Mass}

This section presents an overview of the three steps required to
measure the top quark's mass: 1) generating collisions, 2) selecting
collisions, and 3) measuring mass.

\subsection{Generating Collisions}

\begin{figure}[htb]
\includegraphics[width=3in]{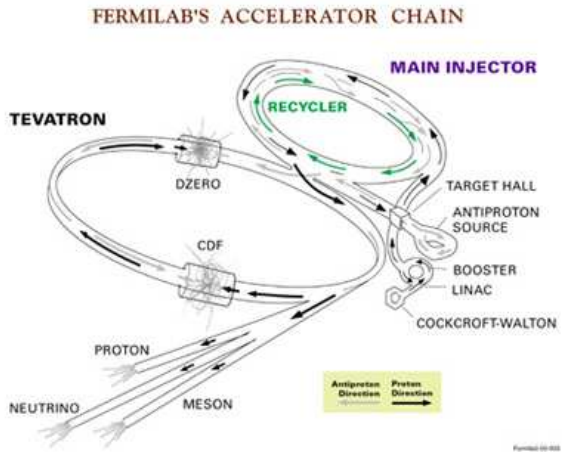}
\caption{ \small
  Accelerator complex at FermiLab, showing the
  chain of lower energy accelerators used to prime the Tevatron, the
  world's highest energy accelerator, which collides protons and
  anti-protons at two points (CDF and DZERO) in the ring.}
\vspace{-0.2cm}
\label{fig:fnal}
\end{figure}

To provide enough energy to produce massive exotic particles such as
the top quark, one must accelerate and annihilate lighter particles
and their anti-particles.  The Tevatron collider at FermiLab
accelerates protons and anti-protons to a center-of-mass energy of
1.96 tera electron-volts, the highest controlled energy collisions
ever achieved.  Figure~\ref{fig:fnal} shows the accelerator complex,
which includes a series of smaller accelerators that seed the final
4-mile Tevatron ring.

Every hour, the Tevatron collider produces approximately $10^{10}$
collisions, the vast majority of which do not produce top quarks. When
produced, the rare top quark cannot be directly observed, as it decays
into a series of lighter particles, called \emph{decay products}, in
approximately $10^{-23}$ seconds.  These decay products can be observed
via multiple layers of detectors~\cite{CDF} that surround the point of
collision and measure the decay products' direction and energy.

\subsection{Selecting Collisions}
\label{sec:select}

\begin{figure}[htb]
\vspace{-0.4cm}
\begin{center}
  \includegraphics[width=2.3in]{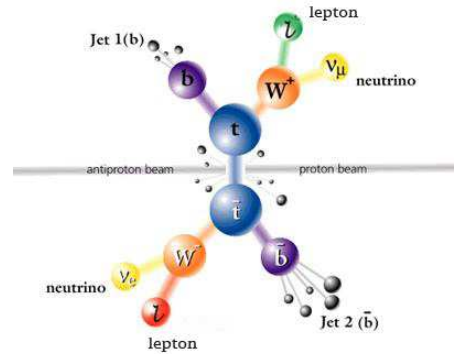}
\end{center}
\caption{ \small 
  Production and decay of top quark pairs, which decay almost
  immediately to a pair of $W$ particles and bottom quarks. $W$
  particles decay to leptons and their accompanying neutrinos while
  bottom quarks decay to jets of lower energy particles.}
\vspace{-0.4cm}
\label{fig:ttbar}
\end{figure}

Most background collisions are removed during a \emph{pre-selection}
phase, which discards all collisions that do not display the top
quark's characteristic signature.  This signature, which emerges from
the top quark's decay products (see Figure~\ref{fig:ttbar}), consists
of two leptons, two jets caused by bottom quarks, and an energy
imbalance caused by missing neutrinos, which escape undetected.

Unfortunately, this signature is not unique to top
quarks~\cite{dilepton}.  After pre-selection, 83\% of the data
sample is expected to consist of backgrounds that mimic the top quark's signature.
In particular, five types of backgrounds may survive pre-selection:
production of 1) two gluons and a $Z$ boson which decays to a pair of 
stable leptons ($ee$ or $\mu\mu$) contributes 71\% of the sample 2) three gluons with a $W$ boson at 6\% 3) two gluons with a $Z$ boson which decays to a pair unstable leptons ($\tau\tau$) at 3\%  4) two gluons and two $W$ bosons at 2\% 5) a $Z$ boson and a $W$ boson at 1\%.

While these backgrounds mimic the top quark's basic signature, they
differ from top quark collisions in more subtle ways, e.g.\ the
distribution of energy in the leptons or jets.  By exploiting these
differences, physicists have devised a heuristic selector that further
prunes the data sample~\cite{dilepton}.  However, more effective
selectors can be generated using machine learning, a process detailed
in Section~\ref{sec:learning}.

\subsection{Measuring Mass}
\label{sec:measuring}

Given a sample of selected collisions, the top quark's mass can be
measured by inferring the likely mass of the observed decay products
in each collision~\cite{dileptonmass}.  The goal in
generating and selecting collisions is to minimize the
uncertainty of this measurement.

\section{Machine Learning for Collision Selection}
\label{sec:learning}

This section describes how collision selection can be performed with
the aid of machine learning.  First, we describe an approach based on
supervised methods.  This approach is standard in the physics
community and serves as a baseline of comparison for the results
presented in Section~\ref{sec:results}.  Second, we present a novel
approach based on stochastic optimization techniques.

\subsection{Supervised Learning for Collision Selection}
\label{sec:supervised}

Training a classifier which separates signal from background with
supervised methods requires a data set of correctly labeled
collisions.  The correct labels for real collisions are not known.
However, physicists have developed a sophisticated simulator of the
collider~\cite{pythia} and detector~\cite{cdfsim} which generates
collisions and models the interaction of their decay products with the
detector.  These collisions are generated using three likely mass values:
165, 175, and 185 giga-electron-volts per speed of light squared
(GeV$/c^2$).

\subsubsection{Training Binary Classifiers}

\begin{figure}[htb]
\begin{center}
\vspace{-0.4cm}
\includegraphics[width=3.5in]{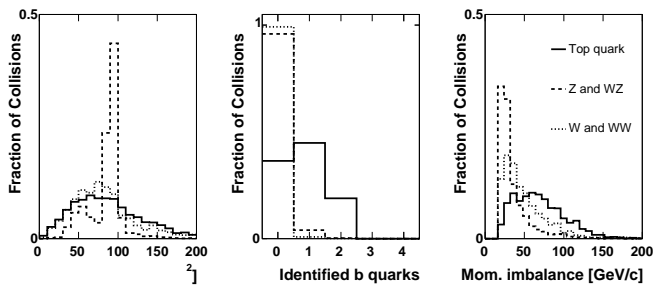}
\caption{ \small 
  The distribution of values in the data set of simulation collisions,
  after pre-selection, for three of the six features.  Left, the mass
  of the two leptons shows a strong peak for $Z$ background.  Center,
  only top quark decays contain real bottom quarks, a powerful
  discriminator when identified.  Right, the momentum imbalance in
  collisions producing top quarks is typically large due to the
  inability to detect neutrinos produced in top quark decay.}
\vspace{-0.4cm}
\label{fig:features}
\end{center}
\end{figure}

In previous research on related collision selection problems, such
examples have served as training sets for binary classifiers
represented as neural networks~\cite{nn2,nn1} or support vector
machines~\cite{svm}.  In this paper, we follow this approach to
produce baseline top quark selectors.  In particular, we train
feed-forward neural networks with six inputs, fourteen hidden nodes,
and one output.  The inputs correspond to six features that describe
each collision: 1) the mass of the system of two leptons, 2) the
number of identified bottom quarks, 3) the imbalance of transverse
momentum, indicating the presence of undetected neutrinos, 4) the
total transverse energy of all decay products, 5) the minimum angle
between a jet and the unbalanced transverse momentum, and 6) the
minimum angle between a jet and a lepton.  Figure~\ref{fig:features}
shows the distribution of values in the data set of simulated
collisions, after pre-selection, for three of these six features.

In training, each collision is labeled 1 if it is signal and 0
otherwise.  In testing, a collision is classified as signal if the
network's output is greater than a threshold $t \in [0,1]$.  Since we
cannot quantify \emph{a priori} the trade-off between precision and
recall, we set $t$ to the value that maximizes classification accuracy
on the training set.  To find this value, we sample the range $[0,1]$
at regular intervals of 0.025, computing the classification accuracy
at each point.

\subsubsection{Training Multi-Class Classifiers}

A potential disadvantage of the binary classification approach is that
it gives all backgrounds the same label.  We hypothesized
that better performance was possible by treating each type of
background as a separate class, yielding a multi-class classification
task instead of a binary one.  We train a set of
\emph{one-against-all}~\cite{one-against-all} classifiers: six binary
classifiers, each of which uses the same network topology described
above and strives to distinguish a given class from all the others.
Hence, when training the $k$th classifier, each collision is labeled 1
if it is in class $k$ and 0 otherwise.  Note that one of these
classifiers, that which distinguishes signal from all five
background classes, is identical to the binary classifier described
above.  In testing, a collision's classification corresponds to the
network with the highest output.

Multi-class classification may outperform binary classification but is
still suboptimal because it only maximizes classification accuracy and
cannot favor harmless background collisions over disruptive ones.
Cost-sensitive learning methods~\cite{cost-sensitive} are designed for
exactly such problems but are inapplicable in collision selection
because the relative costs of misclassification for each class are not
known.  However, stochastic optimization methods, described in the
remainder of this section, are applicable.

\subsection{Optimization Methods for Collision Selection}
\label{sec:stoch}

The supervised approach described above has proven effective at
training collision selectors.  However, it is not ideal because
it assumes that the collision selector with the highest classification
accuracy will result in the lowest mass measurement uncertainty.  In
practice, however, the mass measurement is more sensitive to some
backgrounds than others.  Therefore, selectors that maximize
classification accuracy may perform worse than those that 1) increase
the quantity of signal by tolerating harmless background or 2) reduce
the quantity of signal to eliminate disruptive background.

\subsubsection{Measuring Mass Uncertainty}
To find such selectors, we must define a metric that evaluates the
quality of any given selector.  Ultimately, we want the selector that
produces top quark mass measurements with maximal accuracy and
precision.  However, optimizing collision selectors for accuracy is
unnecessary because the mass measurement is calibrated for accuracy
using simulated collisions, a process known as \emph{bias
  correction}~\cite{dileptonmass}.  Hence, background collisions that
do not lower precision are harmless even if they introduce bias.  The
best selector is that which produces the most precise mass
measurements, regardless of the resulting bias.

The most precise measurements are those with the smallest statistical
uncertainty, which we measure by calculating the standard deviation of
the mass estimates the selector produces on a series of 1000
independent trials at each of the three likely top quark masses.  In
each trial, we randomly select collisions, with replacement, from
the pre-selected training set and feed them to the
selector.\footnote{The number of collisions in each trial was chosen
  to approximately equal the number of collisions produced by the
  Tevatron collider in one year that survive pre-selection.}  The
collisions that survive selection are used to estimate the top quark's
mass, as described in Section~\ref{sec:measuring}. The standard
deviation of these estimates after bias correction reflects the
statistical uncertainty of mass measurements produced by that
selector.

\subsubsection{Optimizing Binary Classifiers}

The simplest way to exploit this new metric is in setting the
threshold $t$ of the binary classifier.  Instead of setting $t$ to
maximize classification accuracy, we set it to minimize mass
measurement uncertainty.  As before, we sample the range $[0,1]$ at
regular intervals of 0.025.  However, at each point, we compute the
mass measurement uncertainty of the resulting selection, not the
classification accuracy.

\subsubsection{Optimizing Multi-Class Classifiers}

Optimizing $t$ could improve performance by effectively balancing the
trade-off between precision and recall.  However, it is still
suboptimal because it treats all background types equally.  A
selector that optimizes the output of the multi-class classifier (made
up of six binary classifiers) could perform much better: by
distinguishing between different background types, it could favor
harmless collisions and discard disruptive ones.

The one-against-all approach to multi-class classification does not
have thresholds to tune.  Nonetheless, its performance can be improved
using stochastic optimization techniques.  Instead of directly using
the classifiers for selection, we use their classifications as input
to a selector trained to minimize mass measurement uncertainty.  This
selector is also a neural network but its structure and weights are
determined by NEAT, a stochastic optimization technique, described
below, that searches for networks that minimize mass measurement
uncertainty.  Since this selector receives as input estimates of the
class of a given collision, it can learn to distinguish between
harmless and disruptive backgrounds.

\subsubsection{Optimizing Selectors Without Supervised Learning}

This paper also investigates a more dramatic departure from the
standard approach to collision selection, one which does not employ
supervised methods at all.  In this approach, the inputs to the NEAT
selector are not the outputs of the one-against-all classifiers but
instead the original six features, described in
Section~\ref{sec:supervised}, that served as inputs to those
classifiers.  As a result, training classifiers is no longer
necessary.  Instead, we treat collison selection purely as an
optimization problem and rely on NEAT to find a selector that
minimizes mass measurement uncertainty.  The remainder of this section
provides a brief overview of the NEAT method and details the two ways
we employ it to train collision selectors.

\subsubsection{Stochastic Optimization With NEAT}

NEAT (NeuroEvolution of Augmenting Topologies)~\cite{neat}, is a
stochastic optimization technique that uses evolutionary computation
to train neural networks.  While many other optimization methods could
be used in its place, we chose NEAT for collision selection because of
its previous empirical success on difficult optimization
tasks~\cite{neat,whiteson:jmlr06}.

In a typical neuroevolutionary system~\cite{yao:ieee99}, the weights
of a neural network are strung together to form an individual genome.
A population of such genomes is then evolved by evaluating each one
and selectively reproducing the fittest individuals through crossover
and mutation.  Most neuroevolutionary systems require the designer to
manually determine the network's topology (i.e.\ how many hidden nodes
there are and how they are connected).  By contrast, NEAT
automatically evolves the topology to fit the given problem.

\begin{figure}[htb]
\begin{center}
\includegraphics[width=2.5in]{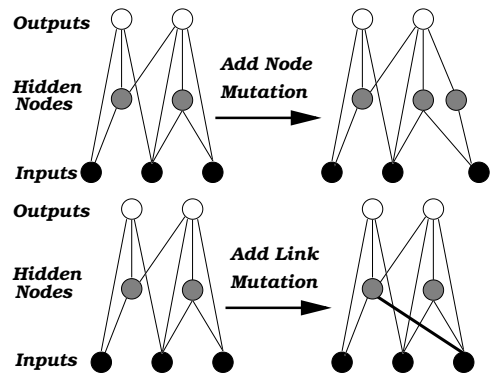}
\caption{ \small
  Examples of NEAT's structural mutation operators.  At top, a new
  hidden node, shown on the right, is added to the network by
  splitting an existing link in two.  At bottom, a new link, shown
  with a thicker black line, is added to connect two existing nodes.}
\label{fig:mutate}
\vspace{-0.6cm}
\end{center}
\end{figure}

NEAT begins with a uniform population of simple networks with no
hidden nodes and inputs connected directly to outputs.  In addition to
standard weight mutations, two special mutation operators
incrementally introduce new structure to the population.
Figure~\ref{fig:mutate} depicts these operators, which add hidden
nodes and links to the network.  Only those structural mutations that
improve performance tend to survive; in this way, NEAT searches
through a minimal number of weight dimensions and finds the
appropriate level of complexity for the problem.

In collision selection, NEAT evolves networks that indicate whether a
given collision should be kept or discarded.  Hence, a network's
inputs describe a single collision, either using the output of the
multi-class classifier or using the original six features, as
described above.  The networks have one output and use the threshold
$t=0.5$ in both training and testing.  It is no longer necessary to
tune $t$ since NEAT evolves networks that are optimized for a fixed
value of $t$.

To operate, NEAT must have a way to measure the fitness of each
network in its population.  In collision selection, the fitness of a
given selector is negatively proportional to the mass measurement
uncertainty, computed as described above.  NEAT favors for
reproduction those networks that minimize this uncertainty.

Finding a good selection in this manner is challenging in part because
of the size of the search space.  The set of possible selections is
the power set of the collisions.  Hence, given $n$ collisions, there
are $2^n$ possible selections.  Nonetheless, directly searching for
selectors that minimize mass measurement uncertainty yields much
better performance than maximizing classification accuracy, as the
results in the following section confirm.

\section{Results}
\label{sec:results}

To assess the efficacy of the methods presented in this paper, we
evaluated each one on ten independent runs, using 10,000
simulated collisions.  These runs were conducted using ten-fold cross
validation: in each run, 75\% of the collisions are selected at random
for training and the remaining 25\% reserved for testing.

\subsection{Supervised Learning Results}

\begin{figure}[htb]
\begin{center}
\vspace{-0.8cm}
\includegraphics[width=3in]{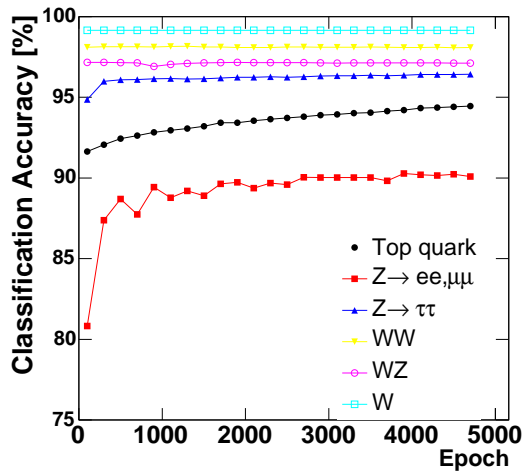}
\vspace{-0.7cm}
\caption{ \small 
  Classification accuracy on training data for the six networks, each
  trained to recognize the signal (top quark) or one of the five
  backgrounds. Accuracy is averaged over all classes, weighted by the expected contributions (see Section 2.2). Networks trained to recognize rare backgrounds therefore have high accuracy.  The signal network is used in the binary  classification case; all six networks are used in the
  \emph{one-against-all} multi-class approach.}
\label{fig:superresults}
\vspace{-0.4cm}
\end{center}
\end{figure}

Figure~\ref{fig:superresults} shows the classification accuracy on
training data for networks trained with backpropagation on simulated
pre-selected collisions, averaged over ten independent runs. As
described in Section~\ref{sec:supervised}, each network is trained to
identify one class of collisions.  Binary classification uses only the
network trained to identify top quark collisions, while multi-class
classification uses all six networks.  The networks had six inputs,
fourteen hidden nodes, one output, and were trained with a learning
rate of $0.001$ and a momentum rate of $0.5$. Accuracy during training
was measured using a threshold $t=0.5$.

On the data reserved for testing, the binary classifier had an average
classification accuracy of $93 \pm 1$\%.  The multi-class classifier
indentified the correct class with an accuracy of $83 \pm 1\%$.  If
the multi-class classifier is not penalized for labelling backgrounds
with the wrong background class, its accuracy improves to $91 \pm
1\%$.  Binary and multi-classifiers give mass measurements with an
average uncertainty of $10.1 \pm 0.4$ GeV$/c^2$ and $10.0 \pm 0.5$
GeV$/c^2$, respectively.

\subsection{Optimization Results}

If the threshold of the binary classifier is selected to minimize mass
measurement uncertainty instead of maximizing classification accuracy,
the resulting selectors allow for mass measurements with substantially
better average uncertainty: $9.1 \pm 0.4$ GeV$/c^2$.

\begin{figure}
\begin{center}
\vspace{-0.1cm}
\includegraphics[width=3.5in]{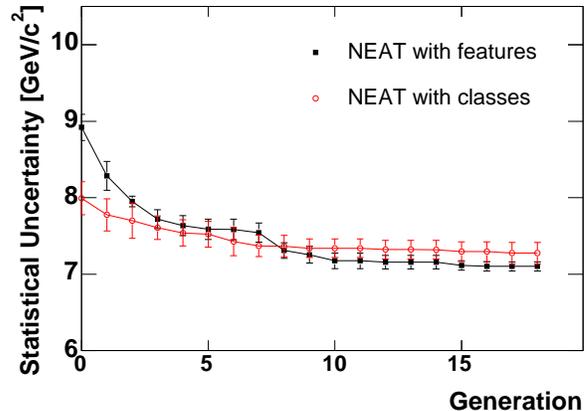}
\vspace{-0.7cm}
\caption{ \small 
  Mass uncertainty in training data for the best network in each
  generation trained with NEAT.}
\vspace{-0.6cm}
\label{fig:neatresults}
\end{center}
\end{figure}

Using NEAT to perform stochastic optimization yields even more precise
measurements.  Figure~\ref{fig:neatresults} shows mass uncertainty on
the training set for the best network in each generation trained with
NEAT.  It compares the performance of NEAT optimizing multi-class
classifiers to its performance optimizing directly on the features,
without the help of supervised methods.  The results are averaged over
ten runs for each method.  In testing, the average mass uncertainty of
the final generation champions was $7.3 \pm 0.3$ GeV$/c^2$ and $7.1
\pm 0.2$ GeV$/c^2$ for the two approaches, respectively.

\begin{figure}[htb]
\begin{center}
\includegraphics[width=2.5in]{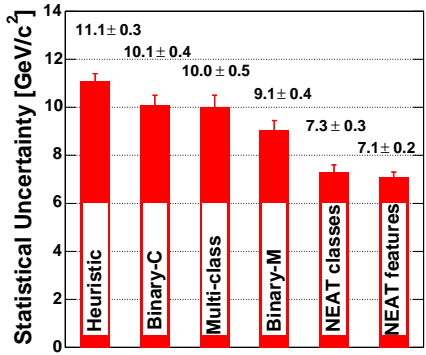}
\vspace{-0.7cm}
\caption{ \small 
  Average mass measurement uncertainty on testing data for the
  heuristic selector, binary classifiers with $t$ optimized for
  classification accuracy (Binary-C), or mass uncertainty (Binary-M),
  multi-class classifiers (Multi-class), NEAT with multi-class
  classifiers as inputs (NEAT classes), and NEAT with the original
  features as inputs (NEAT features).}
\vspace{-0.5cm}
\label{fig:allcomp}
\end{center}
\end{figure}

Figure~\ref{fig:allcomp} summarizes the performance on testing
data of all the machine learning methods we employed and compares it
to the performance of the heuristic selector designed manually by
physicists.  Student's t-tests confirm with $>98\%$
confidence the statistical significance of the differences between 1)
the heuristic selector and each learning method, 2) each supervised
method and each optimization method, and 3) the optimized binary
classifier and each NEAT method. 

\section{Discussion}

The results presented above confirm the conclusion of earlier
work~\cite{nn2,nn1,svm} that machine learning methods can
substantially outperform heuristic collision selectors.  However,
previous results demonstrated only that learned selectors had higher
classification accuracy, while these results directly verify that they 
improve the precision of mass measurements.

Furthermore, these results confirm the advantage of treating collision
selection as an optimization problem rather than a supervised learning
one.  Even the simplest optimization strategy, which tunes the
threshold of a binary classifier, yields significantly more precise
mass measurements than either purely supervised approach.  Using NEAT
to directly search for effective selectors performs even better,
yielding 29\% smaller mass uncertainty than the supervised approach.
Obtaining comparable precision using supervised selectors would
require 66\% more collisions, costing tens of millions of dollars and
hundreds of person-years.

Surprisingly, the multi-class supervised learning approach performs
only as well as the binary approach, which suggests that knowing the
type of a particular background is not helpful in distinguishing it
from signal.  However, multi-class classification performs better than
binary classification when serving as a starting point for
optimization.  This result makes sense since its six outputs form a
richer description of each collision than the binary approach,
which leaves only a single threshold to optimize.  However, the
performance of NEAT given the original features instead of classifier
outputs suggests that supervised methods are unnecessary in this task.
Even without their aid, NEAT is able to quickly discover collision
selectors that yield highly precise mass measurements.

This novel approach to collision selection directly aids the
progress of high energy physics, since the NEAT selectors described in
this paper are currently in use at FermiLab for selecting collisions
from real data collected with the Tevatron collider. Hence,
stochastically optimized collision selectors contribute substantially
to our knowledge of the top quark's mass and our understanding of the
larger questions upon which it sheds light.

\section{Future Work}

In addition to those tested in this paper, many other machine learning
techniques that could aid top quark collision selection.  Optimization
methods like hill climbing, simulated annealing, or other evolutionary
methods could be used instead of NEAT.  In addition, recently
developed methods for structured prediction~\cite{structured} may
improve the performance of supervised methods by allowing them to
minimize arbitrary cost functions like mass measurement uncertainty.

Furthermore, top quark mass measurement is only one of many potential
applications of machine learning techniques to high energy physics.
For example, current theories of particle physics require the
existence of a not-yet-observed particle, the Higgs boson, which gives
mass to other particles through its interactions.  Observation of the
Higgs is one of the primary goals of the Tevatron and its successor,
the Large Hadron Collider near Geneva, Switzerland~\cite{science}.
Extracting the subtle signals of the Higgs boson's decay will require
effective collision selectors.  Hence, we hope to apply stochastic
optimization or other machine learning techniques to aid this search.

\section{Conclusion}

This paper presents a new approach to training collision selectors for
high energy physics.  Rather than relying on supervised methods to
train classifiers that separate signal from background, this approach
uses stochastic optimization methods to directly search for selectors
that minimize the statistical uncertainty in the resulting mass
measurements.  Empirical results on multiple independent trials
confirm that, while supervised approaches outperform heuristic
selectors, stochastically optimized selectors substantially outperform
them both.

\footnotesize

\bibliographystyle{named}
\bibliography{ijcai}

\end{document}